\documentclass[apj,numberedappendix,tighten]{emulateapj}
\usepackage{amsmath}

\def\***#1{***\textbf{\textsf{#1}}***}

\begin{document}

\shorttitle{QSO Structure Function and Variability Mechanism}
\shortauthors{A.~Voevodkin}

\title{Structure Function and Variability Mechanism of Quasars from SDSS
  Stripe 82}

\author{Alexey~Voevodkin\altaffilmark{1,2}}

\altaffiltext{1}{Space Research Institute (IKI), Profsoyuznaya 84/32,
  Moscow, Russia, 117997 }

\altaffiltext{2}{Los Alamos National Laboratory, MS-D466, Los Alamos, NM, 87545}

\begin{abstract}
  Theoretical predictions for the ensemble quasar structure function are
  tested using multi-epoch observations of Stripe 82 collected by the Sloan
  Digital Sky Survey.  We reanalyze the entire available volume of the
  $g$-band imaging data using difference image photometry and build high
  quality light curves for 7562 spectroscopically confirmed quasars.  Our
  structure function includes $\sim4.8\times10^6$ pairs of measurements and
  covers a wide range of time lags between 3 days and 6.9 years in the
  quasar rest frame. A broken power-law fit to this data shows the presence
  of two slopes $\alpha_1=0.33$ and $\alpha_2=0.79$ with the break at
  $\sim42$ days.  The structure function compiled using only flux increases
  is slightly lower than that for variations of the opposite sign, revealing
  a slight asymmetry between the leading and trailing edge of a typical
  flare. The reality of these features is confirmed with monte-carlo
  simulations.  We give simple interpretation of the results in the frames
  of existing theoretical models.

\end{abstract}



\section{Introduction}
\label{sec:intro}

Several decades after the discovery of quasars, the physical origin of their
variability is still not fully understood. An unambiguous explanation of the
intrinsic variations in quasar emission will likely provide the key to
understanding the energy supply of quasars. The most frequently invoked
model connects variability of quasars to instabilities arising in the
accretion disk around a super-massive black hole
(e.g.~\citealt{1998ApJ...504..671K,2006ApJ...642...87P}). In the starburst
model quasar light curves are a superposition of very frequent supernova
explosions in the host galaxy
(e.g.~\citealt{1992MNRAS.255..713T,1996MNRAS.282.1191C,1997MNRAS.286..271A}).
Gravitational microlensing by compact bodies within the host---such as
normal stars or certain candidates for dark matter---also produces
observable flux changes
(e.g.~\citealt{2002MNRAS.329...76H,2007A&A...462..581H}), but their
contribution to the observed level of variability remains uncertain.  The
models can be distinguished based on their predictions for the dependence of
the slope and the amplitude of the structure function on the time-scale of
variability in the quasar rest frame.  Recently, individual quasar light
curves have been successfully modeled as a damped random walk (DRW) process
(\citealt{2009ApJ...698..895K}, \citealt{2010ApJ...708..927K},
\citealt{2010ApJ...721.1014M}).  While the method allows one to study the
structure function of individual objects as a function of black hole mass,
quasar luminosity, wavelength of observation
(\citealt{2010ApJ...721.1014M}), it also fixes the slope of the power
spectrum in the high frequency regime at $-2$ for every quasar.  Moreover,
the DRW model excludes the possibility that the structure function of
positive flux variations differs from the one for negative changes. In this
paper we rely on more traditional statistics using ensemble structure
functions.

A recent influx of survey data on variability of tens of thousands of QSOs
is fueling numerous observational studies and calls for a new look at the
proposed scenarios.  The most recent papers favor the disk instability model
(\citealt{2009ApJ...696.1241B}, \citealt{2005AJ....129..615D},
\citealt{2004ApJ...601..692V}) with the exception of
\cite{2002MNRAS.329...76H} who find evidence in support of the microlensing
model.  Here, we re-analyze the repeated $g$-band imaging of Stripe 82
collected by the SDSS using difference image photometry and build an
ensemble structure function over a wide range of time-scales in order to
test the theoretical predictions.  Our results show that in their present
form none of the considered models explains the behavior of the structure
function on all time-scales. Instead, a hybrid model including both a
starburst and disk instabilities naturally follows from the data.

The paper is organized as follows. In Section~\ref{sec:data} we describe the
data and a method of extracting light curves. Basic properties of the
structure function are given in Section~\ref{sec:sf}.  In
Section~\ref{sec:sf_ensemble} we build the structure function for the
ensemble of quasars.  We discuss the results in Section~\ref{sec:discussion}
and a summary is given in Section~\ref{sec:summary}.


\section{Data analysis}
\label{sec:data}

\subsection{Multi-epoch SDSS Imaging in Stripe 82}
\label{sec:imaging}

For the purpose of this study we reanalyze all available $g$-band drift-scan
images of the SDSS Stripe 82 in the final SDSS data release, hereafter DR7
\citep{2009ApJS..182..543A}.  Quasar light curves were obtained using
difference image photometry \citep{1998ApJ...503..325A,2000A&AS..144..363A}
and a slightly modified version of the photometric software employed by the
OGLE project \citep{2000AcA....50..421W}.  The original motivation for
difference imaging approach was to facilitate a sensitive search for
strongly lensed quasars~\citep{2006ApJ...637L..73K,2009ApJ...698..428L}.
But the application of the method also returns high quality relative light
curves of all sources in a given data set, which we use to study the
variability of known quasars.

\begin{figure}
  \includegraphics[width=\columnwidth]{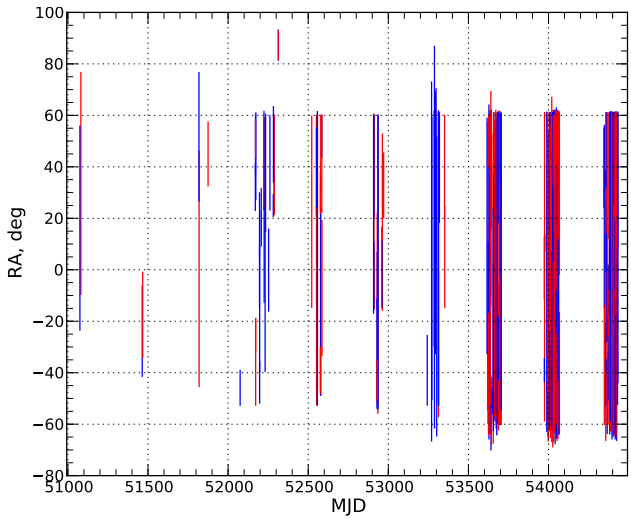}
  \figcaption{Spatial and temporal coverage of the SDSS Stripe 82 imaging
    database. North runs are shown in blue and South runs in red. Note a
    much better time sampling in the later half of the survey.
 \label{fig:geom}}
\end{figure}

The full database of Stripe 82 contains 303 runs covering nearly 300~deg$^2$
of sky along the celestial equator ($-60^\circ\le\rm{RA}\le60^\circ$ and
$-1.25^\circ\le \rm{Decl} \le1.25^\circ$).  The area was scanned repeatedly
during 10 years with some regions observed more than 80 times
\citep{2009ApJS..182..543A}.  The SDSS imaging system consists of 30 CCD
detectors organized in 6 raws (camcols), where a single CCD in each row is
taking data in one of the 5 filters $ugriz$. Since there are gaps between
camcols, the width of the major stripe is split into North and South strips
covered by independent scans. The SDSS database of Stripe
82 contains 162 North and 141 South runs. The spatial and temporal coverage
of all runs is shown in Figure~\ref{fig:geom}, where the blue and red lines
correspond to the North and South runs. Each run
is divided into $2048\times1489$ pix sections called fpC frames with a
$2049\times128$ pix margin on both ends \citep{2002AJ....123..485S}.  The
locations of the beginning and the end of each scan vary from one
observation to another.  This arrangement is not compatible with subtracting
multiple images of the same area, so the runs must be ``glued'' back
together and then subdivided using consistent field boundaries.  The
rebinning starts from RA$_{\rm{min}}$, the first location covered by at
least 20 runs (determined separately for North and South runs). The
rearranged frames of $2048\times1489$ pix are written as FITS files and the
process continues untill we reach RA$_{\rm{max}}$, beyond which fewer than
20 runs are available.

Some of the data in Stripe 82 were obtained in non-photometric conditions
and in relatively poor seeing (DR7). The performance of the difference image
analysis method is sensitive to the quality of the images, and therefore our
next step is to reject images with poor seeing, high background, and
significant background gradients. A small number of images taken toward the
ends of the run have strong background gradients due to the Sun or Moon
rising or setting.  Occasionally, a problem with the bias level subtraction
caused a jump in the count level between the left and right parts of the
image corresponding to individual amplifiers on the CCD chip. Such images
were removed from the analysis. A good quality reference image was prepared
for each field using 10 images with the best overall seeing, background, and
transparency. This is done with the help of the SExtractor
program~\citep{1996A&AS..117..393B} by requiring a median $FWHM\le1.5"$ for
point sources (${\rm CLASS\_STAR}\ge0.7$) and rejecting background maps with
obvious problems. The fields with fewer than 20 remaining images were
discarded.

\subsection{Difference Image Photometry of SDSS Quasars}
\label{sec:photometry}

Interpolation of images to the same pixel grid was accomplished using
approximately 50 stars identified on all images of the field. The time base
line of the Stripe 82 data set approaches a decade. We found that a
significant fraction of stars shows a detectable proper motion producing
time-variable residuals in the shape of a dipole in the difference images
(cf. \citealt{2009ApJ...698..428L}). In order to limit the impact of such
problems on the quality of the final photometry, we selected the reference
stars with the proper motion $\mu\le100$ mas yr$^{-1}$ from the
\cite{2008MNRAS.386..887B} catalog in the overlap area (about 50\%). The
spatial variability of the PSF matching kernel was computed using a
third-order polynomial fit to model coefficients and a second-order
polynomial for the differential background.  Again, slow stars from the
catalog of \cite{2008MNRAS.386..887B} were preferentially used to compute
the solution.  The DIA light curve is computed by adding difference fluxes
measured with the DIA package to the flux in the reference image. Object
magnitudes were converted to standard SDSS $g$ magnitudes by adding a median
offset between stars identified on the reference image and corresponding
objects in the SDSS catalog.

\begin{figure}
  \includegraphics[width=\columnwidth]{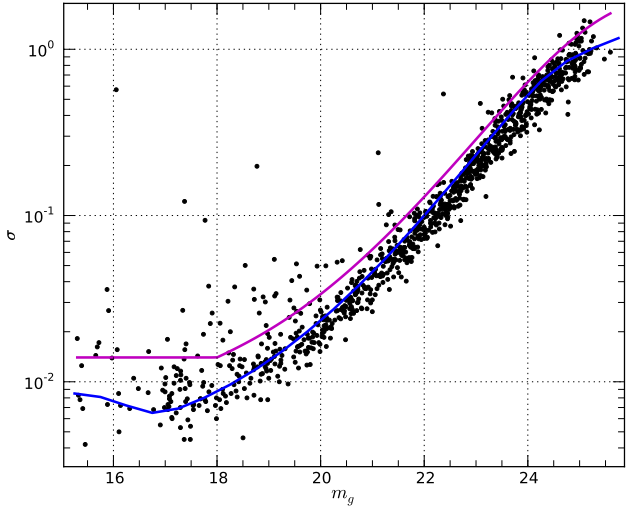}
  \figcaption{Variability diagram in the $g$ band for a typical field in
    SDSS Stripe 82.  The magnitude scatter of all measurements for a given
    source is plotted against the baseline magnitude (median). The
    photometry is based on our reanalysis of the SDSS scans using difference
    image photometry. Black points correspond to the sources identified by
    SExtaractor.  Sources above the magenta line are considered to be
    variable. Blue line shows median variability of the sources from all
    fields.
 \label{fig:qa}}
\end{figure}

The photometric quality is demonstrated in Figure~\ref{fig:qa} showing the
magnitude scatter versus baseline magnitude for a typical field. Here, the
baseline is the median magnitude and the variability of the source is
calculated as the 68-th percentile of absolute deviations from the
median. Black points correspond to the sources identified in that field and
blue line shows median of the variability of the sources from all fields.

In most fields the frame-to-frame accuracy reaches 7 mmag for bright
unsaturated sources. For a significant fraction of stars the photometric
scatter is worse than that and limited by the residuals due to proper
motions.  The photometric uncertainty is 1\% or better for sources with
$m_g\le18$ and around 10\% for sources with $m_g\approx22$. There are 52
quasars with $m_g>22$. Those were excluded from the structure function analysis
Section~\ref{sec:sf_ensemble}) due to large uncertainties in their light curves.
In the next step, we match spectroscopically confirmed quasars from DR7
Quasar Catalog \citep{2010AJ....139.2360S} with our photometric database.
There are 9519 DR7 catalog quasars within the area covered by Stripe 82, and
we can build DIA light curves for 7614 of them. There are several reasons
why we miss a significant number of quasars. After exclusion of poor quality
images many fields are left with fewer than 20 epochs and are removed from
further analysis.  An average 10\% of the image area is lost due to masking
of very bright stars.  Some fields are rejected for lack of stars that can
be used to calibrate light curves. Finally, our pipeline occasionally fails
on a small fraction of images with problems in fpC frame headers.

\begin{figure}
  \includegraphics[width=\columnwidth]{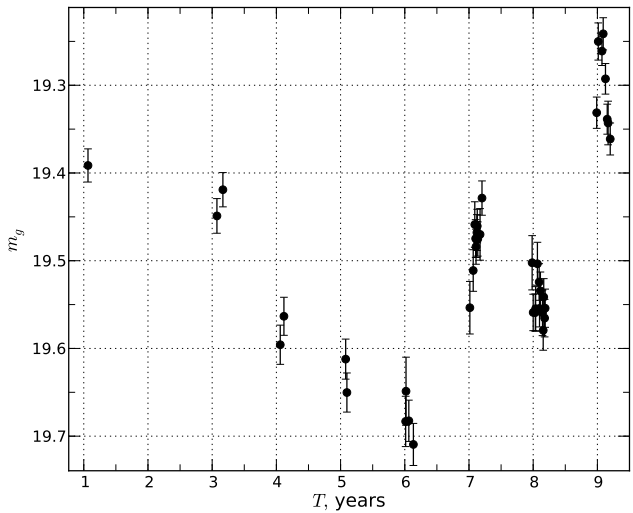}
  \figcaption{Example difference imaging light curve for QSO
    SDSS~$223154.15+004747.5$ at $z=0.8409$.  Time counts from the first
    observation in Stripe 82.
 \label{fig:qso_ex}}
\end{figure}

An example quasar light curve resulting from this procedure is shown in the
Figure~\ref{fig:qso_ex}.  There are 43 observations in the light curve and
time counts from the first epoch in Stripe 82 (run 94, fpC frame 12). The
number of epochs per quasar in our sample varies between 20 and 56 with the
mean of 36.

\section{Structure function}
\label{sec:sf}

The time sampling of our quasar light curves is relatively sparse (see
Figure~\ref{fig:qso_ex}) and insufficient for the use of spectral methods
such as the frequency-power spectrum analysis. Following the common practice
in AGN variability studies, we focus on the information contained in the
structure function.  A number of structure function definitions are in use
differing in relatively minor details that are not important for this
study. We adopt a structure function (hereafter SF) in the form:

\begin{equation}\label{eq:sf}
S(\tau) =
\left\{\frac{1}{N(\tau)}\sum_{i<j}(x_i-x_j)^2\right\}^{1/2},
\end{equation}
\noindent
where $x_i=x(t_i)$ some random quantity at moment $t_i$, the sum is taken
over all measurements for which $t_j-t_i=\tau$ and $N(\tau)$ is the number
of such data points. In short, the structure function is a time dependent
standard deviation of some random process. SF of white noise is flat and
equal to $\sqrt{2\sigma^2}$. Therefore, the SF of a random process is also
flat on time-scales, where the amplitude of the intrinsic variability is
comparable with the measurement errors, as well as on the longest
timescales where the r.m.s. stops changing as the length of the time
interval keeps increasing. In the context of quasar variability the random
quantity is the magnitude (essentially the relative luminosity) $m_i$
measured at time $t_i$ and $\tau$ is taken in the quasar rest frame.  For a
given quasar we can write:
\begin{eqnarray}\nonumber\label{eq:sf2}
S^2(\tau) &=&\langle(m_i-m_j)^2\rangle_\tau \sim
\frac{1}{\bar{f}^2}\langle(\delta f_i-\delta f_j)^2\rangle_\tau \equiv\\
& &\equiv\frac{1}{\bar{L}^2} \langle(\delta L_i-\delta L_j)^2\rangle_\tau=
\frac{2\sigma_\tau^2}{\bar{L}^2},
\end{eqnarray}
\noindent
where we converted to flux units. The flux consists of a stationary
component $\bar{f}$ and a time dependent variable part $\delta f_i$, and
$\sigma_\tau$ is the r.m.s. calculated for points separated by time interval
$\tau$.  From here it is obvious that for two quasars at the same redshift
and with equal $\sigma_{\tau}^2$ the brightest one will have smaller
variability amplitude, $S(\tau)$, at all time scales.

The number of time intervals that can be constructed from an average of 36
epochs for a single quasar is $36\times(36-1)/2 = 630$ and results in a very
noisy structure function\footnote{Structure functions of individual sources
  can be effectively used for identification of quasars (see
  e.g.~\citealt{2010ApJ...714.1194S,2011A&A...530A.122P}).}. Therefore, we
focus our analysis on the ensemble quasar SF, i.e. we assume that the
ensemble variability reflects on average the variability of a single quasar.


\section{The Ensemble Structure Function of Quasars}
\label{sec:sf_ensemble}

\begin{figure}
  %
  %
  \includegraphics[width=\columnwidth]{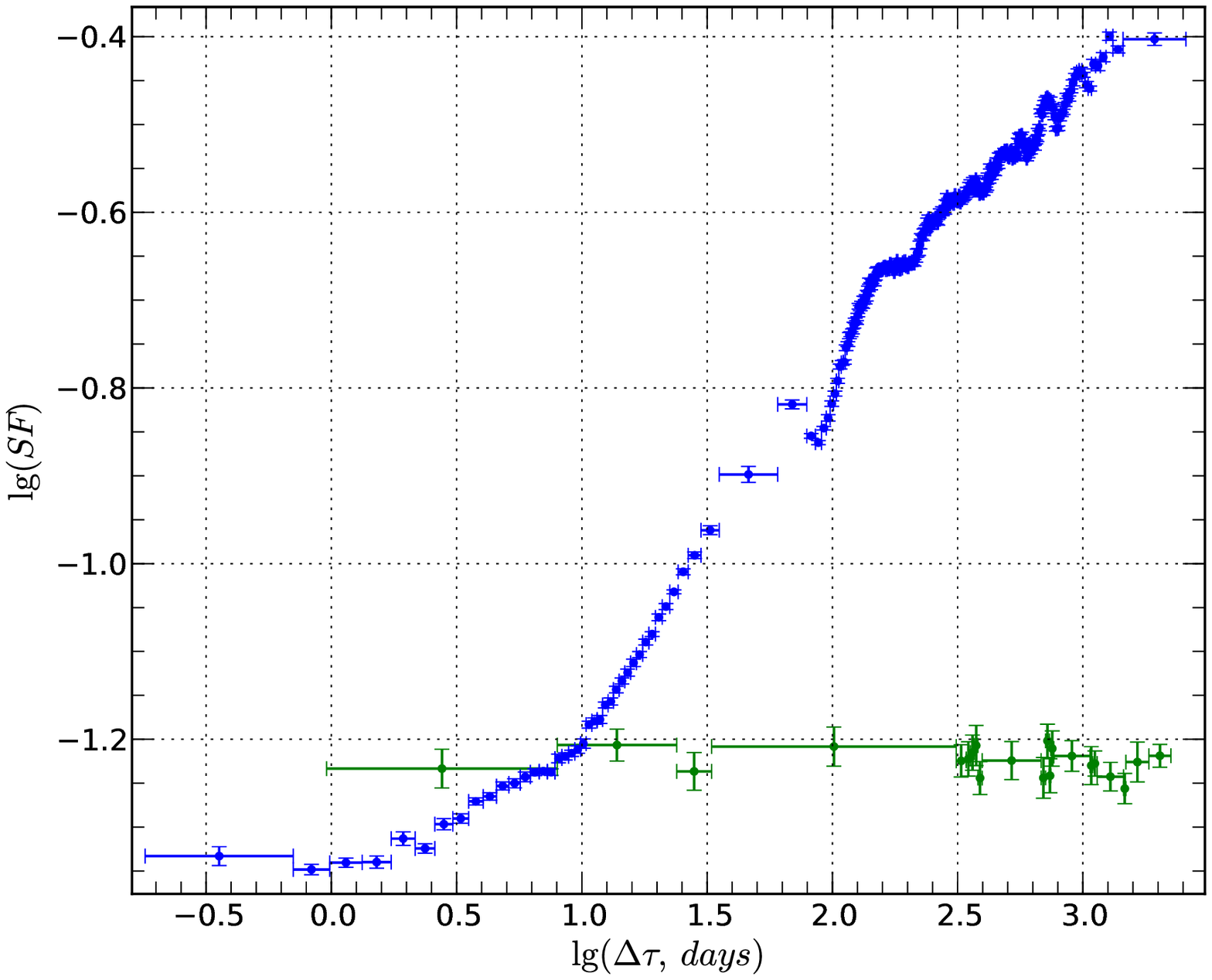}
  \includegraphics[width=\columnwidth]{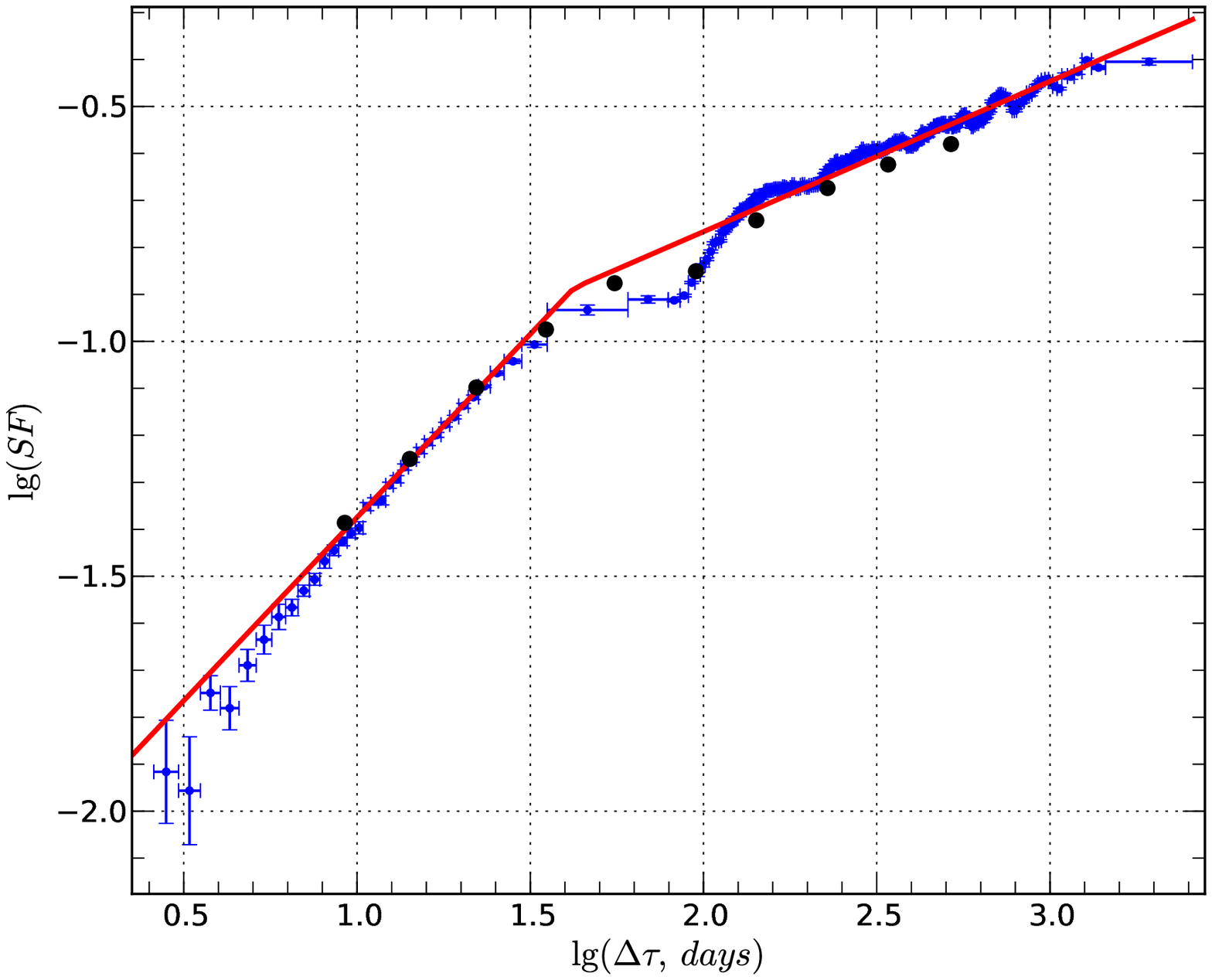}
  \figcaption{ \emph{Top panel:} Raw ensemble structure function of 7562
    quasars from SDSS Stripe 82 (blue) compared to the structure function
    for $\sim1000$ non-variable stars (green). The stars included in this
    comparison fall below the magenta line in Figure~\ref{fig:qa}) is shown
    by green points.
  \emph{Bottom panel:}
    Noise-corrected structure function for an ensemble of quasars (blue) and
    the best fit broken power-law model (red). The oscillations for
    time-scales above $\sim100$ days are an artifact of the time
    sampling. The model offers only an approximate description of the
    structure function over the full extent of the data.
 \label{fig:sf}}
\end{figure}

The final sample contains 7562 light curves. All possible pairs of
measurements within each light curve are collected for a total of 4,796,468
magnitude differences. After reducing the time lag to the quasar rest frame
using spectroscopic redshifts from the SDSS Quasar Catalog
\citep{2010AJ....139.2360S}, the data are binned in time. The bin size was
adjusted to reach 20,000 measurements per bin. The resulting ensemble SF
consists of 239 bins and is shown in Figure~\ref{fig:sf} (blue
points). Error bars were estimated by dividing each timescale bin into 20
smaller bins, evaluating the mean SF in each, and calculating the
r.m.s. scatter of the resulting values (see
also~\cite{2005AJ....129..615D}). 

In order to verify that we detect quasar variability and guard against the
possibility of contamination by unrelated factors, we repeated this
calculation for constant stars. For that purpose we select stars below the
magenta line in Figure~\ref{fig:qa} with $m_g\le22$. The result is shown in
the top panel of Figure~\ref{fig:sf} (green points).  As expected, the SF of
non-variable stars is flat over the entire range of time-scales.  We
conclude that the observed variability of quasars is intrinsic and not
caused by unrecognized measurement errors. The stellar structure function
exceeds the noise level for the quasar sample because the stars selected for
this comparison are fainter on average than the quasars.

The quasar SF in the top panel of Figure~\ref{fig:sf} displays several
slopes and breaks.  It starts flat at $\tau\le2$ days, then at
$2\le\tau\le10$ days gradually transitions to an approximate power-law
increase over $10\le\tau\le40$ days, and finally breaks to a shallower slope
above $\tau\simeq100$. The flat portion of the SF is clearly due to noise
and the transition to the first power-law section is due to SFs of single
quasars with different mean brightness $m_g$ reaching the level of noise at
different $\tau$ (see Equation~\ref{eq:sf2}).
The slopes seen at $10\le\tau\le40$ days and $\tau\ge100$ have a physical origin
and are discussed further below. The time interval between 40 and 100 days
is poorly sampled and difficult to interpret. This is a result of the SDSS
scanning strategy for Stripe 82 that produces relatively few observations
separated by 90--270 days in the observer frame (cf. Figure~\ref{fig:qso_ex})
combined with the mean redshift of the sample $\sim1.5$.

The noise contributes to the SF equally at all timescales and must be
subtracted in order to expose the intrinsic quasar variability. The
noise-corrected SF is shown in the bottom panel of Figure~\ref{fig:sf} (blue
points). The shape of the quasar SF is dominated by two slopes that we
estimate by fitting a broken power law model to the data
\begin{equation}\label{eq:bpl}
  S(\tau) = \left\{ 
  \begin{array}{l l}
    \beta(\tau/\tau_0)^{\alpha_1} & \quad \text{if $\tau\ge \tau_0$}\\
    \beta(\tau/\tau_0)^{\alpha_2} & \quad \text{if $\tau\le \tau_0$}\\
  \end{array} \right ..
\end{equation}
\noindent
We find $\alpha_1=0.33$, $\alpha_2=0.79$, $\tau_0=42.3$ days, and $\beta =
0.13$ (solid red line).  The reduced $\chi^2$ of the fit is far from unity,
i.e. Equation~\ref{eq:bpl} does not provide a complete description of the
data.  However, the main goal here is to detect and estimate the two slopes
in the overall shape of the SF for an ensemble of quasars, and compare those
coefficients with theoretical predictions. We also estimate errors on the
derived parameters using 'jackknife' resampling. Namely, we randomly select
1/3 part of the sample, drop it, and fit ensemble SF built from the rest. We
repeat the procedure 100 times and estimate the error on the given parameter
as r.m.s. of proper distribution. Such estimation gives very small errors
(0.01, 0.02, 3.9 days, and 0.01 for $\alpha_1$, $\alpha_2$, $\tau_0$, and
$\beta$ correspondingly) saying that the shape of ensemble SF is quite
stable to small variations of the sample.

\begin{figure}
  \includegraphics[width=\columnwidth]{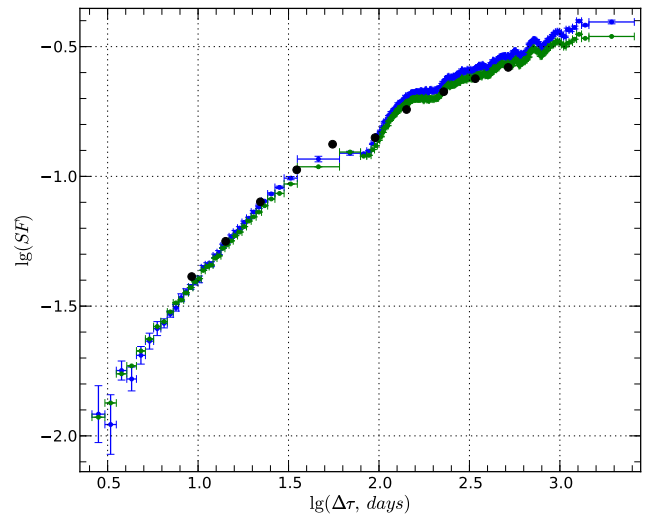}
  \figcaption{Noise-corrected SF$^A$ (blue) and SF$^B$ (green). Black
    points show the $g$-band structure function from
    \cite{2008MNRAS.383.1232W} who used the SF$^B$ estimator. Note a close
    agreement in both the shape and amplitude.
 \label{fig:sf_a_b}}
\end{figure}

For comparison we plot the $g$-band SF from \cite{2008MNRAS.383.1232W}
(black points in the bottom panel of Figure~\ref{fig:sf}). The overall
agreement in shape between the SFs is very good and the discrepancy in
normalization is explained by the different SF type used
in~\cite{2008MNRAS.383.1232W}.  There are two types of SF commonly used in
literature. Following \cite{2009ApJ...696.1241B} we refer to them as SF$^A$
and SF$^B$:
\begin{eqnarray}
\label{eq:sf_a} S^A(\tau)&=&\sqrt{\langle(\Delta m)^2\rangle}\\
\label{eq:sf_b} S^B(\tau)&=&\sqrt{\frac{\pi}{2}\langle |\Delta m|\rangle^2}.
\end{eqnarray}
\noindent
The structure function in the form SF$^B$ was introduced by
\citealt{1996ApJ...463..466D}.  Our noise-corrected SFs and the $g$-band
SF$^B$ from \cite{2008MNRAS.383.1232W} are shown in
Figure~\ref{fig:sf_a_b}. There is a nearly perfect agreement in both shape
and normalization between the B-type SFs derived from the two data sets.

\begin{figure}
  %
  %
  \includegraphics[width=\columnwidth]{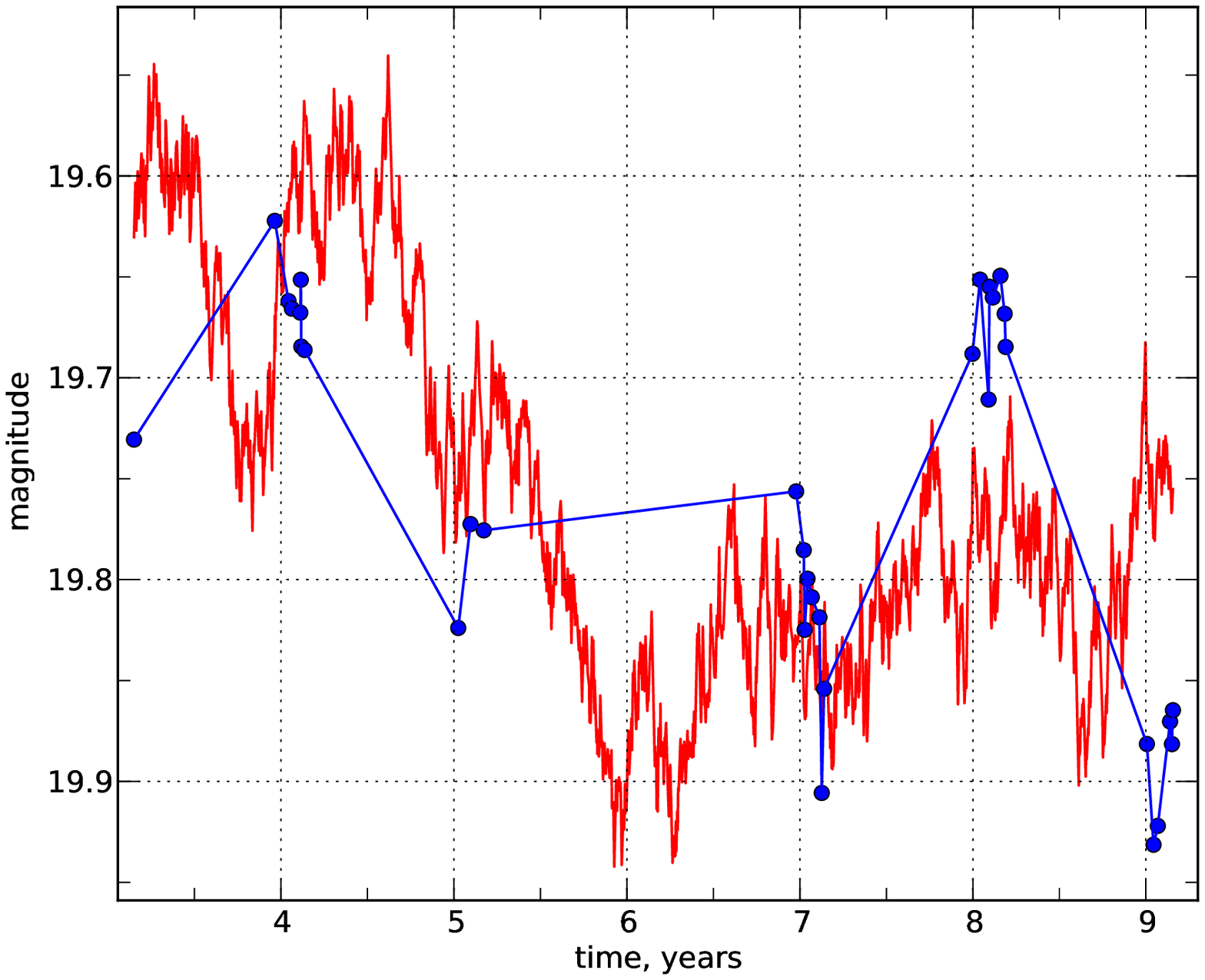}
  \includegraphics[width=\columnwidth]{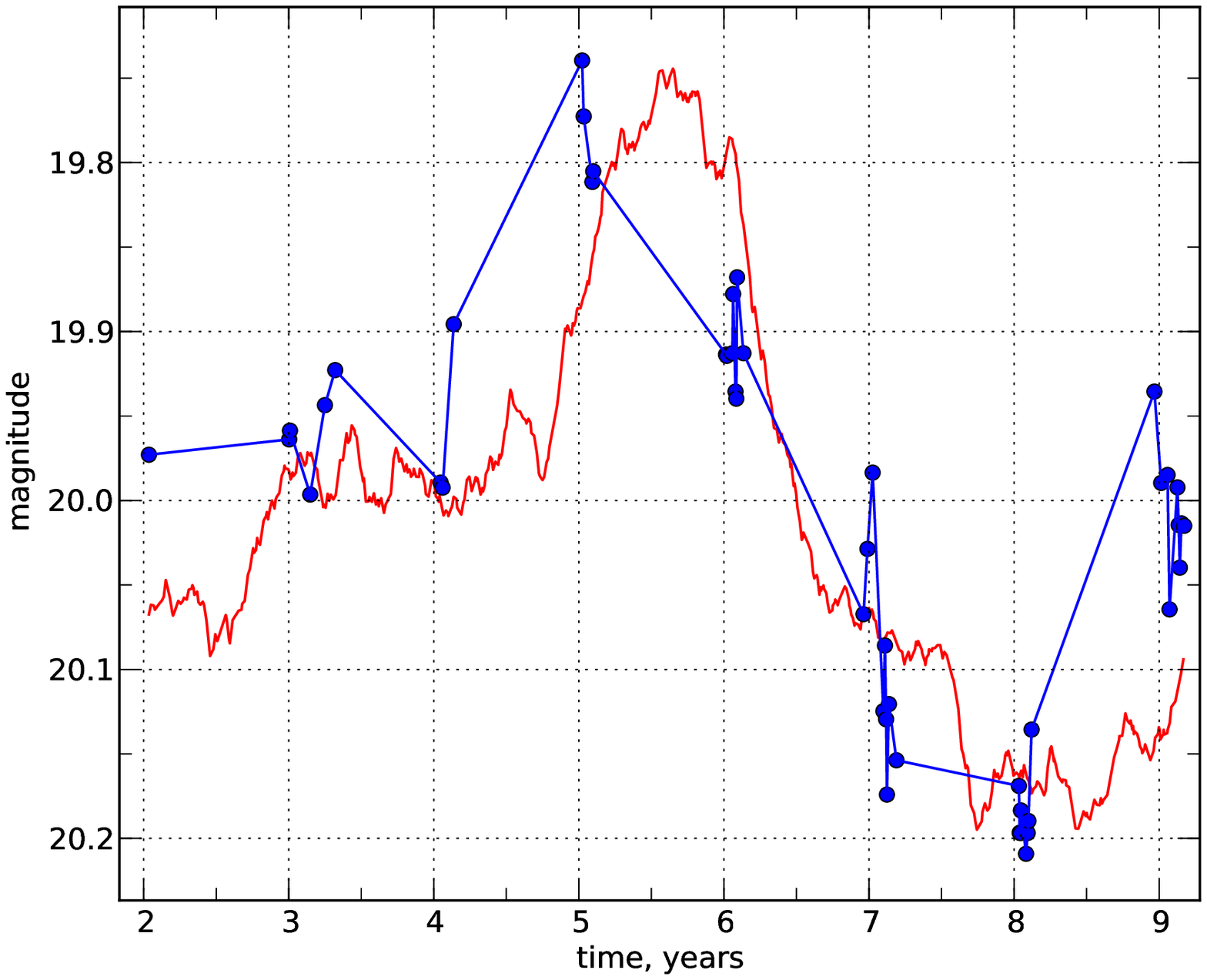}
  \figcaption{Light curve simulations. A random simulated light curve (red)
    is compared with a real light curve of one of the quasars in the sample
    (blue).  The simulated light curves were generated from a power spectrum
    in the form of a single power-law with the slope $-1.66$ (top) and
    $-2.58$ (bottom) corresponding to structure functions in
    Figure~\ref{fig:sim_slopes}.
  \label{fig:sim_lc}}
\end{figure}

The interpretation of results obtained using SF as a statistical tool is not
straightforward. \cite{2010MNRAS.404..931E} show that spurious breaks can
appear in the SF due to sparse time sampling and windowing effects.
Moreover, the slopes derived from SF modeling are typically much more
uncertain than suggested by naive error estimates. Does it mean that the
presence of two slopes in Figure~\ref{fig:sf} is an artifact? To answer this
question and validate our results we perform a set of monte-carlo
simulations.  The goal of this calculation is to verify that for a set of
light curves generated from a single power-law spectrum with a given value
of $\alpha$ and having the time sampling of real quasars, the algorithm of
computing SF does not introduce any significant breaks or biases in the
estimated slope.

The simulation is based on the algorithm of \cite{1995A&A...300..707T} and
reproduces both the correct power spectrum as well as the phase mixture.
The power spectrum in the form $P(f)\propto f^{-(2\alpha+1)}$ corresponds to
SF $S(\tau)\propto\tau^\alpha$. We simulate light curves according to a
prescription in~\cite{2010MNRAS.404..931E}. The simulated time series are
then normalized to have the same r.m.s. and mean as the real light
curves. Examples are shown in the Figure~\ref{fig:sim_lc}.

\begin{figure}
  \includegraphics[width=\columnwidth]{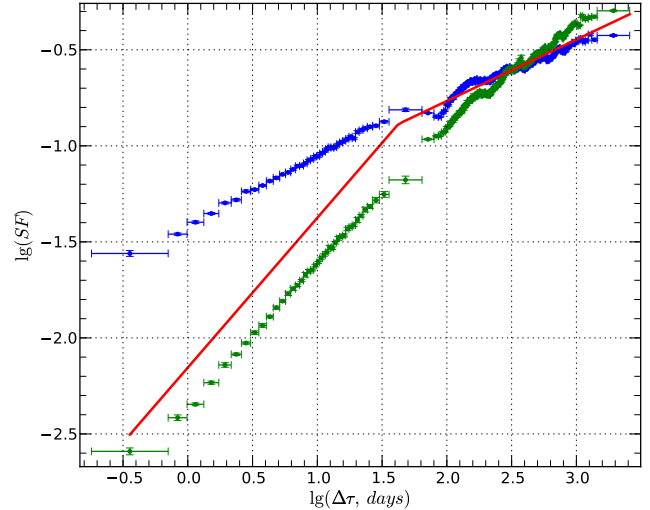}
  \figcaption{
    Confirming the presence of two slopes in the quasar structure
    function. The simulated light curves were generated from a power
    spectrum in the form of a single power-law with the slope $-1.66$ (blue)
    and $-2.58$ (green) corresponding to the upper and lower portions of the
    best fit model (red line) from
    Figure~\ref{fig:sf}.
  \label{fig:sim_slopes}}
\end{figure}

We simulate two sets of light curves with the power-law spectral density
($-1.66$ and $-2.58$), select points on the simulated light curves
corresponding to the time sampling of a real light curve, and apply proper
rest frame correction. In Figure~\ref{fig:sim_slopes} the simulated SFs are
compared to the best fit model for real quasars. The slopes of the simulated
SFs do not exactly match the slope of the model from which they were
generated, i.e. the time sampling slightly biases the measured shape of the
SF. However, none of the simulated SFs is even close to a broken power law
with $\alpha_1 = 0.33$ and $\alpha_2 = 0.79$.  Therefore, the large observed
difference in the SF slope between the short and long timescales is not due
to the time sampling, but rather of a physical origin. Possible
interpretations are discussed in the next section.

\begin{figure}
  \includegraphics[width=\columnwidth]{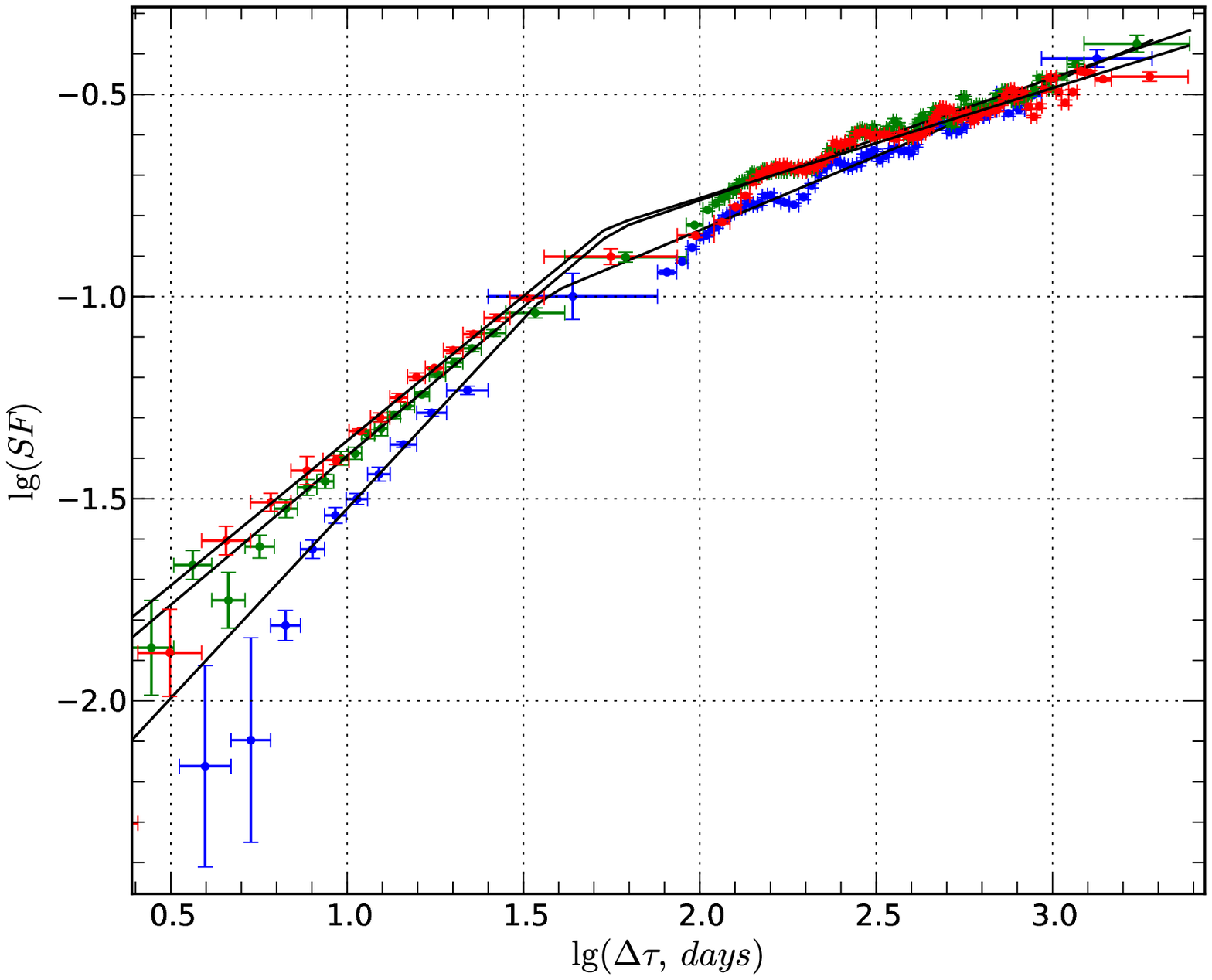}
  \includegraphics[width=\columnwidth]{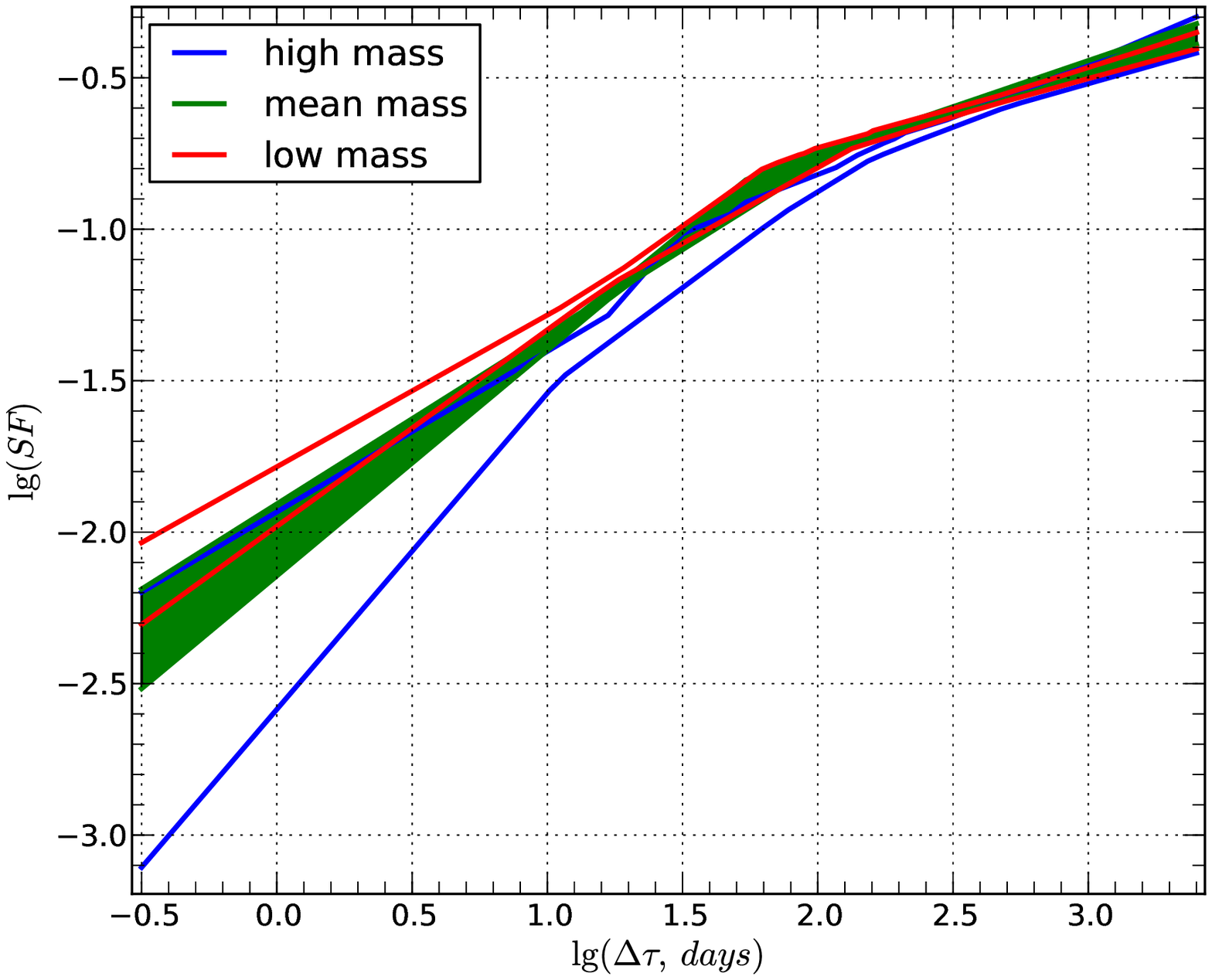}
  \figcaption{ \emph{Top panel:} Ensemble SF for quasars with black hole
    masses lying in low mass range (red points), mean mass range (green
    points ), and high mass range (blue points) (see text for definition of
    ranges). Thin black lines show best fits for each SF.  \emph{Bottom
      panel:} Regions showing the scatter of the best fits obtained during
    'jackknife' resampling. Red lines show the scatter of the low mass range
    SFs best fits, green region and blue lines show the same for mean mass
    and high mass ranges correspondingly.
    \label{fig:sf_mbh}}
\end{figure}

The ensemble SF considered here is built from quasars with different
intrinsic characteristics such as black hole mass or luminosity. Therefore,
it is interesting to check whether two slopes in SF are characteristic for
the hole range of, say, black hole masses or have dependence on mass. Using
black hole mass estimations from~\cite{2008ApJ...680..169S}, we select three
subsamples of quasars with black hole masses in ranges
$\le5\times10^8\,M_\odot$ (1953 sources, low mass range), $5\times10^8\div
1.5\times10^9M_\odot$ (2108 sources, mean mass range),
$\ge1.5\times10^9M_\odot$ (1276 sources, high mass range) and build SF for
each range of masses.  All three SFs show existence of two slopes (see
Figure~\ref{fig:sf_mbh}, top panel). We fit SFs by broken power low model
(Equation~\ref{eq:bpl}). The best fit parameters and formal estimation of
errors by 'jackknife' resampling are given in the Table~\ref{tab:fit_par}. In
order to have better understanding of the uncertainties in the parameters we
also plot regions showing minimal and maximal values of the fits for all
three SFs (see Figure~\ref{fig:sf_mbh}, bottom panel). As it is seen from
the Figure~\ref{fig:sf_mbh}, the SFs slopes tend to increase with the growth
of the mean subsample mass but still have common values. More subtle study
is needed in order to show that the trend in slopes as a function of black
hole mass is real. Here we only conclude that the presence of two slopes in
the ensemble SF is independent on the considered black hole mass range.
  
\begin{table}
  \caption{Best fit parameters}
  \label{tab:fit_par}
  \centering
  \medskip\def\arraystretch{1.15}
  \begin{tabular}{lcccc}
    \hline
    \hline
    Subsample & $\alpha_1$ & $\alpha_2$& $\tau_0$, days & $\beta$\\
    \hline
    low mass  & $0.27\pm0.02$ & $0.72\pm0.04$ & $55.0\pm16.7$ & $0.15\pm0.02$\\
    mean mass & $0.30\pm0.01$ & $0.74\pm0.03$ & $57.3\pm6.6$  & $0.15\pm0.02$\\
    high mass & $0.37\pm0.02$ & $0.94\pm0.09$ & $36.6\pm15.6$ & $0.10\pm0.02$\\
    \hline
  \end{tabular}
\end{table}


\section{Discussion}
\label{sec:discussion}

We consider three processes frequently invoked to explain the variability of
quasars: instabilities in the accretion flow around a supermassive black
hole, supernova explosions related to the starburst phenomenon, and
gravitational microlensing by compact bodies in the host galaxy
\citep{1998ApJ...504..671K,2002MNRAS.329...76H}. SFs of light curves
predicted by each model are characterized by different slopes. The expected
SF slopes are $0.25\pm0.03$, $0.44\pm0.03$, $0.83\pm0.08$ respectively for
variability generated by miscrolensing, disk instability, and starburst
mechanisms \citep{2002MNRAS.329...76H}.  The SF slopes found in
Section~\ref{sec:sf_ensemble} are 0.79 and 0.33 correspondingly for time
lags below and above $42$ days. Taken at face value, the first slope is
consistent with the starburst model, and the second slope falls half way
between the predictions of the disk instability model and the microlensing
model.  Note that SF slopes in the range 0.30--0.35 have been obtained in
other works, but the authors still attributed the variability to disk
instabilities.  \cite{2009ApJ...696.1241B} and \cite{2011A&A...525A..37M}
present an overview of recent results.

\begin{figure}
  \includegraphics[width=\columnwidth]{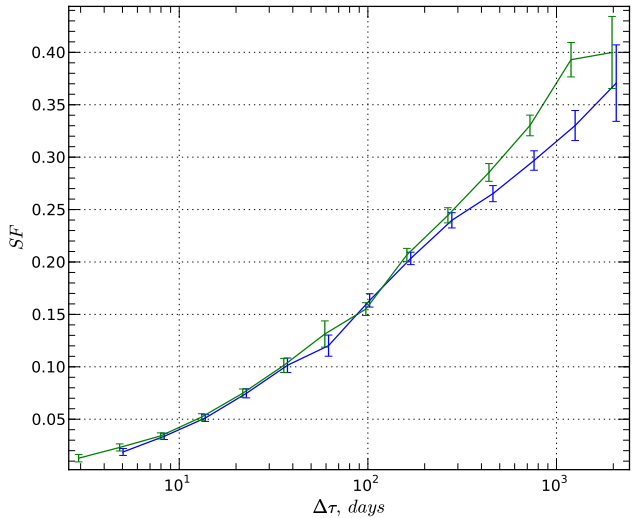}
  \figcaption{Testing the asymmetry in the quasar structure function
    between the leading and trailing branch of a typical flare. Flux
    increases contribute to $S_+$ (blue) and flux drops contribute to $S_-$
    (green).
 \label{fig:sf_plus_minus}}
\end{figure}

Asymmetries in the SF provide an additional test of model predictions.  The
SF that only includes those pairs of measurements for which the flux
increases with time, $S_+$, may be different from $S_-$, the SF
characterizing a decrease in flux. \cite{1998ApJ...504..671K} show that in
starburst models $S_+>S_-$. Disk instabilities produce $S_->S_+$ and
microlensing is symmetric, i.e. $S_-=S_+$ when averaged over sufficiently
long time intervals \citep{2002MNRAS.329...76H}. The difference between
$S_-$ and $S_+$ depends on the parameters of the model and therefore $S_-
\simeq S_+$ is still possible in both the starburst and the disk instability
model \citep{1998ApJ...504..671K}.

Noise-corrected $S_+$ and $S_-$ functions are shown in
Figure~\ref{fig:sf_plus_minus} with blue and green lines respectively. For
long time lags in the range 300--1600 days, where the SF slope is
$\sim0.33$, we have $S_- > S_+$, as predicted by the disk instability model.
The significance of the difference is not very high, but allows us to
exclude the microlensing model. The difference disappears for time-scales
below 100 days, where the slope is consistent with the starburst model.
This may suggest that a typical supernova rate in a starbursts in the host
galaxies of quasars is as high as $\sim100$ yr$^{-1}$ and effectively washes
out any detectable asymmetry (see Fig. 6
in~\cite{1998ApJ...504..671K}). However, the rate should be treated with
caution since it is not derived from direct measurements. Moreover, the
predictions for $S_+$ and $S_-$ were derived from simulations of a single
variability mechanism. The addition of another variability process can wash
out the difference between $S_+$ and $S_-$.  This is an interesting topic
for future research.  The difference in $S_+$ and $S_-$ was also
investigated by \cite{2005AJ....129..615D} and
\cite{2009ApJ...696.1241B}. The former study found tentative evidence that
$S_+ > S_-$ (light curves have fast rise and slow decay as in starburst
model) for 400 days $\le\tau\le$ 1500 days, while in the latter work $S_+
\approx S_-$ over the full range of measurements ($\tau\le1000$ days).  The
explanation of the difference in results is difficult because each paper is
based on entirely different data and requires further research. The
\cite{2005AJ....129..615D} analysis relies on cross survey flux comparisons.

The introduction of the supernova variability mechanism may be redundant as
the shape of the SF can be explained by detailed variability mechanisms
acting in accretion disks. For instance, the PSD of X-ray light curves of
some active galactic nuclei have two (e.g. ~\cite{2007MNRAS.378..649S}) or
more distinct slopes~\citep{2007MNRAS.382..985M}. Considering that the
optical emission tends to follow the X-ray flux with some time delay one may
expect that PSDs of optical light curves will display two slopes. Moreover,
the SF shape found here is in qualitative agreement with the model proposed
in~\cite{2009MNRAS.397.2004A}, where large amplitude optical variations on
the time scales of hundreds of days are attributed to the fluctuations in
accretion rate and short-term (order of days), small amplitude variations
arise due to reprocessing of X-rays.



\section{Summary}
\label{sec:summary}

We applied the image differencing technique to the entire $g$-band imaging
data set of SDSS Stripe 82 and constructed high quality light curves for
7562 spectroscopically identified quasars with no less then 20 epochs. The
variability analysis was performed using the ensemble structure
function. While the shape and normalization of the quasar structure function
derived here are in perfect agreement with the results of
\cite{2008MNRAS.383.1232W} based on an earlier SDSS data release, our
structure function is estimated from a much larger data set and covers a
substantially wider range of time lags from 8 hours to 6.9 years in the
quasar rest frame.

We find that the ensemble SF reveals two distinct power-law slopes with the
break around $42$ days and confirm the presence of these features with
monte-carlo simulations. The presence of two slopes in the ensemble SF is
independent on the black hole mass range of quasars used to build it. The
slopes estimated from a broken pawer-law fit to the data are
$\alpha_1\simeq0.79$ for $\tau\le100$ days and $\alpha_2\simeq0.33$ for
$\tau\ge300$ days.  Using predictions from the theoretical models, the
variability of quasars on times scales $\tau\le100$ days can be explained by
a starburst model. The slope of the structure function on time-scales longer
than about 3 months is rather close to models where variability is explained
by gravitational microlensing.  However, on time-scales $\tau\ge300$ days we
detect a significant asymmetry in the SF characteristic for disk
instabilities that rules out a substantial contribution of microlensing.  It
is also possible to explain the shape of SF without invoking the starburst
model, but using the model, where short term optical variations are caused
by X-rays reprocessing and long therm variations originate from optical
emitting regions of accretion disk~\citep{2009MNRAS.397.2004A}.

The existing SDSS Stripe 82 data offer a great opportunity to better
understand the origin of variability in quasars. The main limiting factor at
present is the lack of detailed theoretical predictions that go beyond the
basic characteristics such as the SF slope and include the detailed shape
and normalization of the SF, as well as asymmetries and color information.

\acknowledgements

AV thanks Przemys{\l}aw~Wo\'zniak for his invaluable contribution to the
preparation of the paper, Eugene Churazov for fruitful discussions on
the statistics of the structure function and acknowledges support from the
research grant RFFI 10-02-01442-a.

Funding for the SDSS and SDSS-II has been provided by the Alfred P. Sloan
Foundation, the Participating Institutions, the National Science Foundation,
the U.S. Department of Energy, the National Aeronautics and Space
Administration, the Japanese Monbukagakusho, the Max Planck Society, and the
Higher Education Funding Council for England. The SDSS Web Site is
http://www.sdss.org/.

The SDSS is managed by the Astrophysical Research Consortium for the
Participating Institutions. The Participating Institutions are the American
Museum of Natural History, Astrophysical Institute Potsdam, University of
Basel, University of Cambridge, Case Western Reserve University, University
of Chicago, Drexel University, Fermilab, the Institute for Advanced Study,
the Japan Participation Group, Johns Hopkins University, the Joint Institute
for Nuclear Astrophysics, the Kavli Institute for Particle Astrophysics and
Cosmology, the Korean Scientist Group, the Chinese Academy of Sciences
(LAMOST), Los Alamos National Laboratory, the Max-Planck-Institute for
Astronomy (MPIA), the Max-Planck-Institute for Astrophysics (MPA), New
Mexico State University, Ohio State University, University of Pittsburgh,
University of Portsmouth, Princeton University, the United States Naval
Observatory, and the University of Washington.


\end{document}